# Stem Cell Transplantation As A Dynamical System: Are Clinical Outcomes Deterministic?

Amir A Toor MD, [1] Jared D Kobulnicky MD, [1] Salman Salman, [1] Catherine H Roberts PhD, [1] Max Jameson-Lee PhD, [1] Jeremy Meier BS, [1] Allison Scalora MS, [1] Nihar Sheth MS, [2] Vishal Koparde PhD, [2] Myrna Serrano PhD, [2] Gregory A Buck PhD, [2] William B Clark MD, [1] John M McCarty MD, [1] Harold Chung MD,[1] Masoud H Manjili PhD, [3] Roy T Sabo PhD,[4] Michael C Neale PhD.[5]

[1] Department of Internal Medicine, Stem Cell Transplant Program, Massey Cancer Center, [2] Center for the Study of Biological Complexity, [3] Department of Microbiology and Immunology, [4] Department of Biostatistics, and the Department of Psychiatry and Statistical Genomics, [5] Virginia Commonwealth University, Richmond, VA 23298

**Correspondence: Amir Ahmed Toor MD,** Associate Professor of Medicine; Stem Cell Transplant Program, Massey Cancer Center, Virginia Commonwealth University, 1200 Marshall Ave, Richmond, Virginia 23298.
Email: atoor@vcu.edu
Phone: 804-628-2389

**Abstract**

Outcomes in stem cell transplantation (SCT) are modeled using probability theory. However the clinical course following SCT appears to demonstrate many characteristics of dynamical systems, especially when outcomes are considered in the context of immune reconstitution. Dynamical systems tend to evolve over time according to mathematically determined rules. Characteristically, the future states of the system are predicated on the states preceding them, and there is sensitivity to initial conditions. In SCT, the interaction between donor T cells and the recipient may be considered as such a system in which, graft source, conditioning and early immunosuppression profoundly influence immune reconstitution over time. This eventually determines clinical outcomes, either the emergence of tolerance or the development of graft versus host disease. In this paper parallels between SCT and dynamical systems are explored and a conceptual framework for developing mathematical models to understand disparate transplant outcomes is proposed.

Stem cell transplantation (SCT) represents a unique immunotherapeutic modality in which donor-derived T cells exert a graft versus host response, which when directed at host-derived malignancy, effects a cure (1, 2). However when this phenomenon extends to normal host tissue, it results in the single most dreaded complication of this procedure, graft versus host disease (GVHD). Over the years more stringent definition of human leukocyte antigen (HLA) identity in donor-recipient pairs (DRP) has diminished the likelihood of GVHD in HLA matched pairs undergoing unrelated donor SCT, (3,4) such that in large patient populations it is seen less frequently. But, take an individual patient - even one with a well-matched sibling donor - and it is entirely impossible to predict whether that individual will develop GVHD, requiring life-long immunosuppression, or become a tolerant chimera, able to come off immunosuppression (5, 6). Aside from the peri-transplant pharmaco-therapeutic interventions, a number of biological factors impact the risk of developing GVHD (7). These include, mismatching of the minor histocompatibility antigens (8), the cytokine milieu (9, 10), and the 'regulatory' immune cell populations (11) in circulation at the time of transplantation.  So despite increasing stringency of HLA matching, a substantial number of patients develop post transplant complications, either related to GVHD or to immunosuppression (infection, relapse), contributing to therapeutic failure as evidenced by the frequent observation of high transplant related mortality following SCT (3, 12, 13). This suggests that outcomes following SCT are inherently stochastic and subject to rules governing probability. So is there some way individual outcomes may be determined following SCT, in other words, is it possible to compute the fate of a transplant recipient?

*Do early conditions affect late outcomes?*

To ascertain this, a quantitative determination of the likelihood of the various post-transplant outcomes would have to be made in different situations. As noted above, HLA matching represents a critical variable in determining survival in transplant recipients. Examining the disparity in clinical outcomes of the patients transplanted using HLA matched and mismatched donors may give an indication of quantitative effect of genetic variation at the MHC locus and the therapeutic adjustment required to overcome that. Over the last decade transplant outcomes observed in patients undergoing alternative donor SCT have steadily improved with relatively minor adjustments to transplant technique. As an example, poor outcomes following umbilical cord blood transplantation (UCBT) in adults were improved by infusing two cord blood units, despite the HLA mismatch between the recipients and the donor cords (14, 15). Graft loss likelihood as well as infection rates declined and no increase in GVHD was observed, even though in the long run only one of the cord blood units would engraft. In a strictly quantitative sense, if not qualitative, the stem cell dose was not significantly altered with the double cord blood infusion when compared with the dose administered using an adult donor, where it was an order of magnitude higher. Similarly, SCT from a haploidentical related donor had been consistently fraught with poor outcomes until the institution of cyclophosphamide infusion on day 3 and 4 following transplant. This has resulted in a marked improvement in survival following SCT with haploidentical donors, even in the absence of T cell depletion (16, 17). In both these examples, interventions early in the transplant course led to a lasting impact on the long-term outcome, with no further intervention beyond the norm. This occurred despite lack of HLA identity, and has led to these mismatched donor sources now being considered viable alternatives if HLA-matched donors are not available. Even when HLA-mismatched unrelated donors are considered, although the transplant risk is higher compared to an HLA-matched donor, with modern conditioning and GVHD prophylaxis regimens, survival and GVHD incidence is relatively similar regardless of whether donors are mismatched at either the allele or antigen level (18, 19). Further, in HLA matched unrelated donors early interventions such as infusion of anti-thymocyte globulin (20, 21) or bortezomib (22) prior to stem cell infusion has resulted in marked impact on long term outcomes. As an example, a small difference in the dose of ATG given during conditioning may have long term effects on the clinical endpoints occurring much later in the course of transplant, presumably by impacting immune reconstitution (23). These examples illustrate the principle that, conditions early on in the course of transplantation are critical in determining long-term outcome, to the extent that they may compensate HLA mismatch. This sensitivity to early conditions is a characteristic of deterministic systems, as opposed to systems governed by randomness.

Further evidence of long-term effects of early conditions comes from examination of immune reconstitution following HLA matched SCT. It has been a consistent observation that early donor derived lymphoid recovery is associated with improved clinical outcomes (Figure 1); less graft loss and relapse, albeit, at the expense of greater GVHD risk (24-29). Conversely, poor donor derived lymphoid recovery either in the form of mixed chimerism or in the terms of low absolute lymphocyte count puts patients at risk for eventual graft loss or relapse, particularly when reduced intensity conditioning regimens are being used (30, 31).

*Are transplant outcomes deterministic?*

Within patients with normal immune recovery there remains an inability to predict whether they will develop alloreactivity or not. This has been explained by the presence of different minor histo-compatibility antigens (mHA) outside the major histocompatibility (MHC) locus. Numerous studies have documented the association of various specific mHA, or groups of mHA, with alloreactivity. However, when whole exome sequences of the SCT donors and recipients were compared, identifying *all* the single nucleotide polymorphisms in a unique DRP, and thus the potential mHA between them, an extensive library of thousands of potential variant mHA was seen in HLA matched pairs, making it unlikely that GVHD occurrence can be explained on the basis of *histo-incompatibility* alone (32). The large burden of minor histo-incompatibility implies that the likelihood of alloreactivity manifesting clinically may be determined by the degree of allo-antigen presentation at the time of transplant, which in turn is determined by the degree of tissue injury and inflammation. The immunosuppressive milieu at the time the initial interaction between T cells and antigen presenting cells occurs then becomes a critical factor in determining tolerance or alloreactivity emerging. The principle at hand appears to be that, all donor recipient pairs will have immunogenic potential for alloreactivity, and in most instances very early on in the course of SCT they will be propelled on a path to certain clinical outcomes (tolerance vs. GVHD vs. graft loss), in a deterministic fashion.

Further support for determinism comes from immune recovery following SCT, which follows predictable kinetics in terms of the order in which various immune cell subsets reconstitute. Commonly, NK cell recovery is prompt, within a few weeks of transplantation followed by cytotoxic T cell recovery, with B cells and helper T cells lagging significantly, especially in patients undergoing T cell depletion. When T cell subsets emerging following SCT are examined with respect to the T cell receptor β (TRB) repertoire complexity, oligoclonal expansion has been observed, which over time recovers back to a more normal repertoire. Importantly, when studied using next generation sequencing (NGS), the T cell repertoire is not disordered, rather, it has a fractal ordering with respect to gene segment usage, which may be described mathematically (33). Fractals describe the geometry of many objects in nature, and are characterized by self-similarity over different scales of measurement. In the human T cell repertoire, proportionality in magnitude is maintained across scales of measurement, when T cell clonal frequency is examined in terms of TRB, variable, diversity and joining gene segment usage. This suggests that a fractal model may be appropriate to describe immune reconstitution following SCT, strengthening the argument for SCT outcomes being deterministic. Given its immunoablative nature, SCT provides a good opportunity to examine the recovery kinetics of T cells, which appear to be influenced by the donor type and the conditions at the time of cell infusion, i.e. use of T cell depletion, or immuno-modulators. Thus, even though the rate of T cell reconstitution may vary in individuals, quantitatively it may be defined mathematically, and this illustrates the principle that T cell repertoire reconstitution kinetics follows a deterministic course.

## Stem cell transplants as dynamical systems

Considering these principles, sensitivity to early conditions, which in a complex background of antigenic diversity leads to divergent outcomes, arrived at by computable immune response; one may postulate that SCT when viewed in individual donor-recipient pairs is an example of a *dynamical system*. In other words, when followed over the course of time each future state of the system (transplant DRP) is dependent upon the state immediately preceding it, rather than being a random occurrence. Dynamical systems evolve over time, and this evolution is modeled by differential equations. These systems may be precisely predictable, as in an accelerating object, where depending on the physical characteristics of the object, one would get the anticipated acceleration every time energy is applied. On the other hand outcomes in certain dynamical systems, may be more difficult to precisely predict, in other words *chaotic*, as in the case of weather, where a *complex system* influenced by a large number of variables, demonstrates disparate outcomes because it's evolution over time is extremely sensitive to initial conditions. Thus even though the behavior of chaotic systems is governed by mathematically described rules, as the system goes through successive *iterations* over time, the eventual outcomes in different individuals diverge exponentially as a function of time. This occurs because minor differences in initial conditions get magnified with the passage of time as the system evolves in each individual. The important concept to recognize in these systems is that if the initial conditions can be faithfully reproduced, chaotic systems will generally have similar outcomes each time, however even very small fluctuations in these conditions sends the system down a different trajectory to an altogether different outcome *state* in different individuals or instances. Further, all the possible potential outcomes, or *states*, constitute the *phase space* of that system, and generally individual systems tend towards a limited number of states, mathematical entities termed '*attractors*' (34-37).

Clearly SCT does not follow our first dynamical system model, since despite the most well designed conditioning regimens and stringently selected donors, outcomes in individual patients are highly variable. Laws of probability can give the odds of a certain outcome, but cannot chart the course an individual will follow after SCT. Further, between genomic variation between donor and recipient, donor-derived T cell repertoire, recipient cytokine milieu and microbiome as well as pathogen exposure, the number of variables to consider is much too great to expect linear, predictable behavior. Therefore, in view of the above discussion, dynamical system modeling of SCT is necessary to understand disparate outcomes, particularly when sensitivity to early conditions is borne in mind. To accomplish this differential equations describing the kinetics of T cell clonal reconstitution over time following SCT, and relating them to the eventual development of either GVHD or tolerance (relapse) may be developed to explore this idea. In such a model, the GVHD risk will depend upon the cumulative effect of the proliferating T cell clones in a deterministic fashion, rather than in a probabilistic manner.

So is it important to distinguish between stochastic or deterministic outcomes following SCT? There is an important difference between the two models: in the former, GVHD or tolerance or relapse, would develop randomly, without any underlying rule or principle being followed. In dynamical systems, however there is an underlying set of rules that the system follows, and if the early conditions can be precisely replicated, the outcomes in different individuals will be more likely to be similar as the system evolves. Acknowledging the difficulty of replicating initial conditions in SCT models precisely, knowledge of the rules at work in SCT would nevertheless permit measured as opposed to empiric therapeutic interventions, such as by more accurate titration and timing of cellular and pharmacotherapy to achieve desired clinical outcomes, within the limits of the system.

*Evidence for the dynamical system model*

What evidence exists that SCT represents a dynamical system? The most telling evidence is the sensitivity to early conditions; consider that in an HLA-matched SCT, minor histocompatibility antigens (mHA) are a constant presence; these are there on the first day of transplant, as they are one year later when the donor-derived T cells are fully reconstituted. Yet bortezomib or ATG or cyclophosphamide given during conditioning may result in the development of tolerance in certain individuals, which in most instances does not break even after withdrawal of immunosuppression. Regardless of the mechanism of how this is achieved, the average impact on the individual system in this instance is that the donor T cells are propelled towards a specific outcome - tolerance - which in this case would be analogous to an attractor, an endpoint to which a chaotic dynamical system tends as it evolves over time. GVHD on the other hand would represent an alternative attractor in the system. An example of this is seen in lymphoid (T cell and NK cell) recovery during the first two months following SCT, influencing eventual outcomes following SCT, whether they be survival, relapse or GVHD (Figure 1) (14, 15, 38). The system 'trajectory', or output may be modified by an intervention; such as donor lymphocyte infusion (DLI) or intensification of immunosuppression to treat GVHD, but left to itself it will tend towards one of the 'attractors'.

Additional support for a chaotic model of SCT comes from the fractal organization of T cell repertoire. Chaotic systems may be represented geometrically as fractals, which demonstrate iterating patterns across scales of magnitude. T cell clonal frequency when considered in terms of T cell receptor β, variable, joining and diversity gene segment usage has a fractal organization. This results in a complex repertoire comprised of thousands of T cell clones, which when examined in terms of clonal frequency, also follow a Power distribution, characteristic of self-similarity across scales of measurement, at all levels of TRB clonal definition (33). Further, when the genomic variability between donors and recipients is considered (32), and translated into putative mHA, the binding affinity of the resulting peptides to the relevant HLA demonstrates a non-linear, Power law distribution, reminiscent of the T cell clonal distribution (Figure 2) (39). Therefore, one may postulate that the driving force behind T cell reconstitution after SCT is the spectrum of binding affinities of recipient mHA (and pathogen) peptides with the relevant HLA in the individual transplant DRP, encountered by donor T cells in the recipient. This is possibly the case, as is evident in a comparison of the peptide-HLA binding affinity distribution and T cell clonal frequency distribution from two different studies performed by our group (Figure 2C and 2D). In this model, depending on the initial conditions following SCT (T cell dose infused + cytokine milieu + pharmacotherapy) specific donor T cell clones will proliferate or decline in a deterministic manner. The antigen presentation in the very beginning will result in either alloreactive or pathogen specific T cell clones proliferating over time, and will eventually determine the clinical outcome, either tolerance or GVHD (Figure 3).

*Modeling T cell clonal expansion in SCT*

One model that may describe the cellular immune recovery following SCT is the logistic model of growth first described by Verhulst in 1838 to explain population dynamics. Logistic growth is described by an equation of the form:

$$x_{t+1} = r\, x_t\, (1 - x_t)$$

In this equation, population size ($N$) at discrete intervals of time ($t$, $t+1$, $t+2...$) is represented as a ratio, $x$, of the possible maximum population size at a much later time $t_n$ (carrying capacity, $K$).

This ratio ($x = N/K$), at any given time in the evolution of a population (for example, $x_{t+1}$) is always determined by the population ratio from an earlier time ($x_t$). In this *iterating equation* the term, $r$ represents the maximum intrinsic growth rate of the population and is called the '*driving parameter*' (36, 40). This relationship has several implications; first, as the population (in this case clonal frequency of individual T cell clones) grows over time, its size at some final time will depend on both the size of the starting population at $t_0$, and the value of $r$. Second, after an initial period of exponential growth, the growth rate slows down asymptotically because the term ($1-x_t$) becomes smaller as the population increases. Third, as the value of $r$ increases, the variance observed in $x$ over time increases, eventually behaving in a chaotic manner. This is depicted in the Logistic Map, where the values $x$ takes on in the long-term, are plotted against $r$ (http://mathworld.wolfram.com/LogisticMap.html). This demonstrates that while the value of $x$ diminishes to zero over time when $r$ is <1, a steady increase in the value of $x$ is observed as $r$ goes from 1 to 3; at $r$ >3 and <3.5, the system may take on two different sets of values of $x$ (bifurcation), consistent with a population oscillating between two extremes; and finally, at $r$ >3.5 the system behaves chaotically with large and unpredictable variation in the value of $x$ (and $N$) over time. Despite this seemingly chaotic behavior, however, if the logistic map is examined at ever-smaller scales (higher decimal place values of $r$) its bifurcation patterns of $x$ seen in the larger map are reproduced in a self-similar manner at each scale of magnification, revealing hidden structure in the distribution of $x$ with each increment in $r$, in other words, fractal organization. If individual T cell clones are considered as unique populations, this provides a plausible explanation for the fractal T cell repertoire observed in SCT recipients.

Extrapolating this model to individual T cell clones followed over time after SCT, one would observe very different growth rates depending on the parameter $r$ governing the growth of each clone. And even though the proliferation of the T cell clones follows deterministic rules, chaotic behavior (if $r$ is high enough) means that, though the eventual clonal frequency of unique T cell clones will be difficult to predict precisely, the overall repertoire will demonstrate underlying order, as was observed in the fractal ordering of the T cell repertoire. Further, the independence of $x$ from $N$ in the logistic equation means that as the Logistic function iterates for each clone, relative proportionality is maintained between T cell clonal populations as they vary over time, resulting in the scale invariance characteristic of fractal geometry. In such a model the individual T cell clones may differ in their frequency by orders of magnitude (41), however this can be addressed by employing a more complete and complex model of growth, such as the Gompertz curve, which by taking $Log\ x$, accounts for the logarithmic nature of growth in biological systems. A potential additional advantage of this is that, it may describe sigmoid population growth more accurately than the Logistic growth curves while also explaining chaotic growth behavior (42, 43).

This hypothesis is supported by the lymphocyte reconstitution kinetics observed in patients undergoing MRD and URD SCT using an immunoablative reduced intensity conditioning (Clinicaltrial.gov identifier: NCT00709592). In this regimen, Thymoglobulin (either 5 or 7.5 mg/kg) was administered in divided doses from day -9 thru -7 followed by 4.5 Gray fractionated total body irradiation. GVHD prophylaxis was with tacrolimus (day -2 to approximately day 120) and mycophenolate mofetil (MMF; day 0-30) (Figure 4A). In patients who achieved hematopoietic engraftment, lymphocyte reconstitution could be plotted as a function of time using a sigmoidal growth curve described by the logistic equation. Most patients had two discernable periods of exponential increase in absolute lymphocyte counts, one following engraftment (Figure 4B), and another period following cessation of MMF (Figure 4C). These growth periods were followed by a plateau with relatively stable lymphocyte counts in the absence of clinical events, until withdrawal of tacrolimus when greater variability was observed (Figure 4D). Patients developing complications of therapy such as relapse, viral (CMV or EBV) reactivation, and GVHD requiring immunosuppression, such as corticosteroid therapy, had significant departures from the sigmoidal curve, as did patients with delayed engraftment kinetics (Figure 4E). It should be recognized that this data represents *all* the lymphocyte subsets, such as NK, T and B cells. The initial phase is likely largely derived from NK cell recovery while the secondary phase most likely represents T cell reconstitution. The latter postulate is consistent with the finding that the day 60 *donor-derived CD3+* cell count is predictive of clinical outcomes (25). It is logical that total lymphocyte counts would reflect the kinetics of T cell clonal proliferation, and that this should be governed by the same principle that describes population dynamics in general, i.e. it is a logistic function of time. Therefore, lymphocyte count growth rate ($r$) observed in our cohort may represent an average of the growth rate of thousands of T cell clones following stem cell transplantation. Thus a simple logistic growth model of T cell reconstitution may be developed which would explain GVHD occurrence as an exponential increase in alloreactive T cells, when immune reconstitution is considered as an iterative process over time (Figure 3).

In SCT, $r$ for each T cell clone may depend upon multiple variables, including the antigen-HLA specificity of the T cell receptor (Figure 2A), the immunosuppressive regimen being used and the cytokine milieu during the period of growth as well as the proportion of regulatory T cell clones. Further, it may vary as immunosuppression is withdrawn following SCT (Figure 4A) or inflammatory states, such as CMV reactivation or GVHD develop leading to increasingly chaotic behavior of the T cell clones. On the other hand, rate of change of $x$ will depend not only on the infused T cells, but will vary as hematopoietic precursors engraft and depending on thymic integrity differentiate into immune cell populations. It is important to recognize that in this dynamical immune reconstitution model the chaotic behavior is occurring at the level of the *individual T cell clones*, and while individual clones may demonstrate marked variance in their frequency over time following transplant, it is their cumulative effect, which results in GVHD or tolerance.  If a large number of mHA directed T cell clones proliferate, then the consequence would be GVHD. Conversely, if non-mHA directed T cell clones dominate, tolerance ensues with immune reconstitution (Figure 3). In such a hypothetical system, the total T cell count trends reflect the average effects of this phenomenon, and the clinical outcomes are an effect of this chaotic expansion of individual T cell clones, with GVHD and tolerance serving as the attractors. It may be postulated that the restoration of a more 'complete' fractal structure will result in optimal clinical outcome. Studies demonstrating oligoclonal T cell expansion in patients with GVHD or relapse demonstrate the validity of this hypothesis (44-47). This concept is testable by measuring the fractal dimension of the post-transplant T cell repertoire by high throughput sequencing (39). Therefore, if one can account for the complexity at hand in SCT, perhaps by using next generation sequencing to study the antigenic variance between donors and recipients, as well as T cell clonotypes following SCT, it will likely be well described as a chaotic dynamical system. Serial high-throughput sequencing of TRB may allow plotting of the T cell clonal frequency as it evolves over time following SCT, resulting in a plot which would yield a fractal surface expanding over time, as individual T cell clones vary and new clones emerge. Such analyses will likely be valuable in distinguishing different prognostic groups of patients on the basis of post-transplant T cell repertoire reconstitution.

Using the dynamical system model, one might monitor the rate of change, in other words $r$, for the dominant T cell clones following SCT and correlate this with clinical manifestations to determine if this is associated with outcomes. Accurate mathematical modeling of the dynamical evolution of T cell repertoire following SCT would allow for *measured,* and *earlier* therapeutic intervention in the event of either GVHD or inadequate immune recovery or residual disease. For example, this may be an intervention using more intense or prolonged post-transplant immunosuppression, for patients with rapid rate of change of $x$ or those with a high value of $r$, to reduce the risk of GVHD by reducing the chaotic behavior if a large number of T cell clones have a high $r$. Conversely, DLI may be similarly used when the opposite conditions prevail. Further, knowledge of the critical time periods when exponential T cell clonal growth occurs with different transplant regimens will allow optimal timing of interventions such as vaccination. For instance, vaccination may be best given before the exponential rise in T cell numbers begins to maximize utilization of cytokines, and minimize potential competition between T cell clones. This concept is utilized in lympho-depleting chemotherapy regimens used as a part of adoptive immunotherapy (48). Therefore considering SCT as a dynamical system rather than a stochastic one, would allow logic based patient management and quantitative trial designs minimizing empiric interventions.  As such, therapy may be designed for individual patients based on a systematic and personalized approach, instead of relying on population-based outcomes derived from probabilistic study designs. In essence, the development of accurate mathematical models that account for the key variables influencing transplant outcomes has the potential to improve clinical outcomes following SCT, making SCT an even better example of personalized medicine than it already is.

**Acknowledgements.**

The authors gratefully acknowledge Ms. Cheryl Jacocks-Terrell for her excellent help in preparing this manuscript.  This study was supported in part by grant funding from the VCU-Massey Cancer Center, from Virginia's Commonwealth Health Research Board, (Grant #236-11-13) and by Sanofi-Aventis.

**Conflict of Interest**: Dr. Toor has received research support from Sanofi-Aventis manufacturers of Thymoglobulin

**Figure 1.** Early lymphoid recovery influences clinical outcomes following allogeneic SCT. Absolute lymphocyte count (ALC) at 1 month predicts survival. As 1-month ALC increased by one-tenth, the odds of survival increased by over 3% (HR = 3.25; 95% CI: 1.59-6.62; P= 0.001). Similarly, as 1-month ALC increased by one-tenth, the odds of relapse decreased by over 3% (HR = 0.33; 95% CI: 0.16-0.66; P= 0.002) *not shown*. Adapted from, 24.

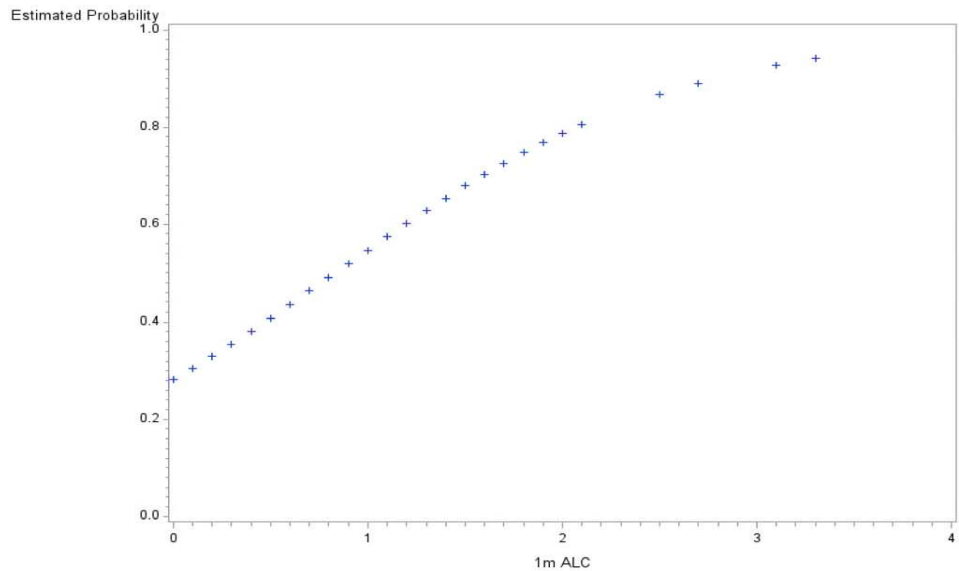

**Figure 2.** (A) Model depicting the relationship between donor T cell clonal frequency and recipient mHA-HLA binding affinity. (B) Postulated association between peptide-HLA binding affinity and T cell clonal frequency distribution. (C) T cell clonal frequency distribution [1] and (D) the values of reciprocal of IC50 [2] (mHA-HLA binding affinity estimate) for mHA-HLA in a single DRP. Both parameters follow a Power law distribution, suggesting that peptide-HLA affinity spectrum has an important role in determining T cell repertoire.

**A.**

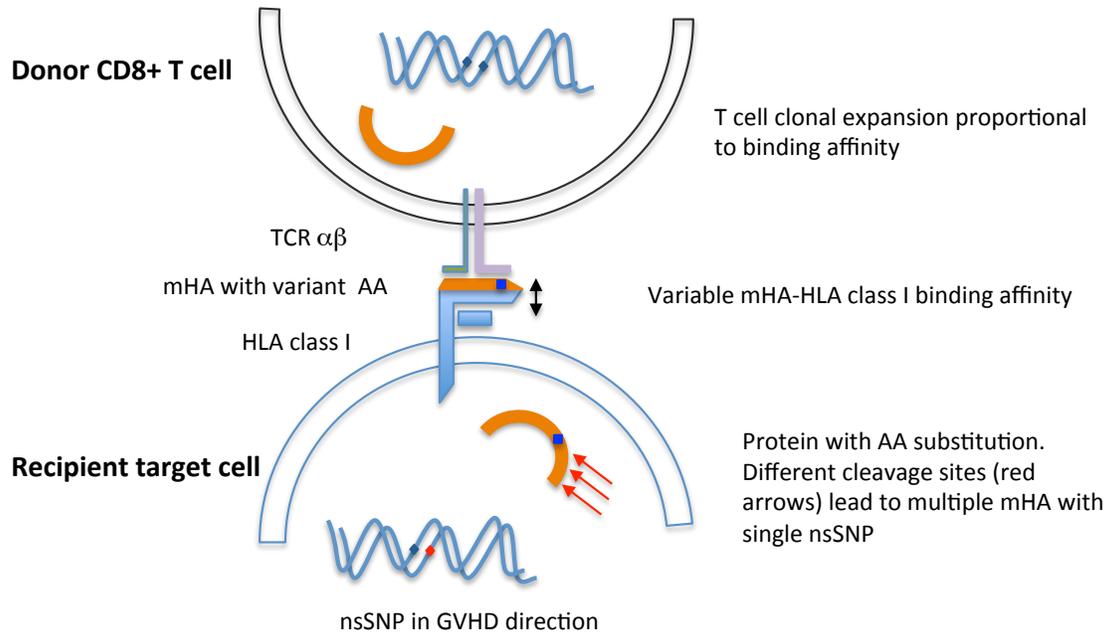

**Donor CD8+ T cell**

T cell clonal expansion proportional to binding affinity

TCR αβ

mHA with variant AA

Variable mHA-HLA class I binding affinity

HLA class I

**Recipient target cell**

Protein with AA substitution. Different cleavage sites (red arrows) lead to multiple mHA with single nsSNP

nsSNP in GVHD direction

**B.**

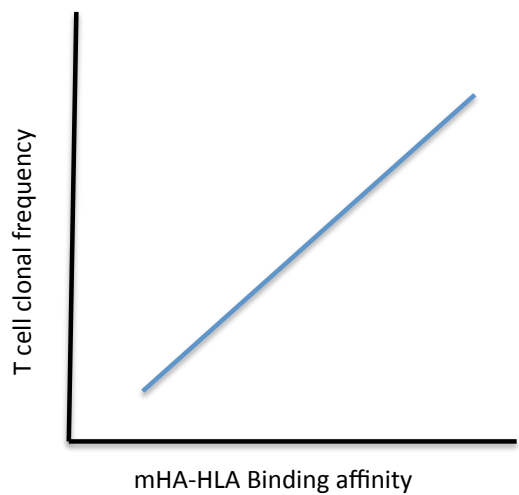

T cell clonal frequency

mHA-HLA Binding affinity

**C**.

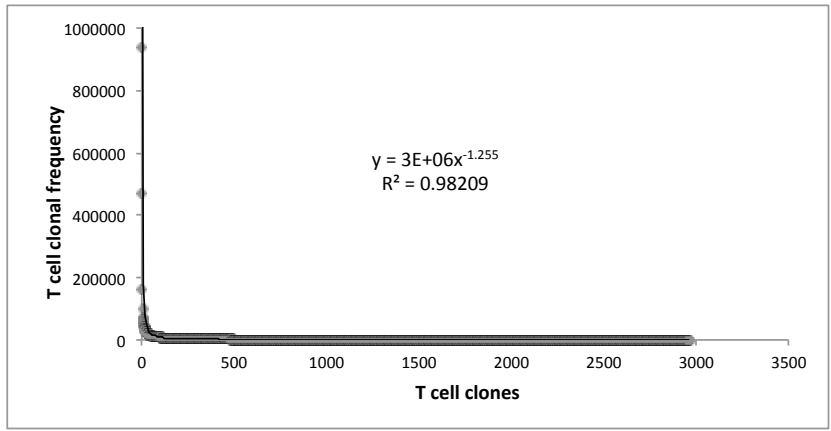

**D**.

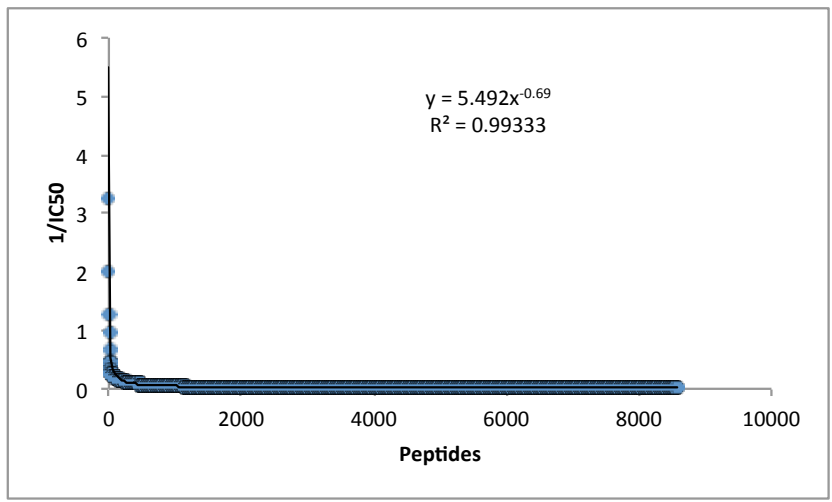

Figure legend- (1) T cell clonal frequency measured on day 100 post SCT, by high throughput sequencing of T cell receptor β, cDNA obtained from CD3+ cells, given in copy number of unique clones and arranged in descending order with a cutoff at <100 copies (2) 1/IC50 of mHA-HLA (estimate of the binding affinity), determined by whole exome sequencing to identify nsSNPs between donor and recipient in the GVH direction, followed by *in silico* determination of the IC50 of the resulting mHA-HLA complexes.

**Figure 3.** Modeling Stem Cell Transplantation as a dynamical system. Iterative expansion of donor T cells clones over time in the presence of an alloreactivity potential, modulated by the degree of antigen presentation. In (A) cells colored red, represent alloreactive T cell clones, and green cells, other non-alloreactive T cell clones. Alloantigen exposure or lack thereof in the first few days of transplant results in minor early differences in the repopulating T cell clones, which over time results in an exponential expansion of corresponding T cell clones. Different phases of cellular proliferation are labeled 1, 2, 3 and 4 in the schema and extrapolated to the plots depicting absolute lymphocyte counts from two patients following SCT (B). These plots show a bi-logistic growth pattern, reflecting initial engraftment and cessation of mycophenolate mofetil following SCT.

A.

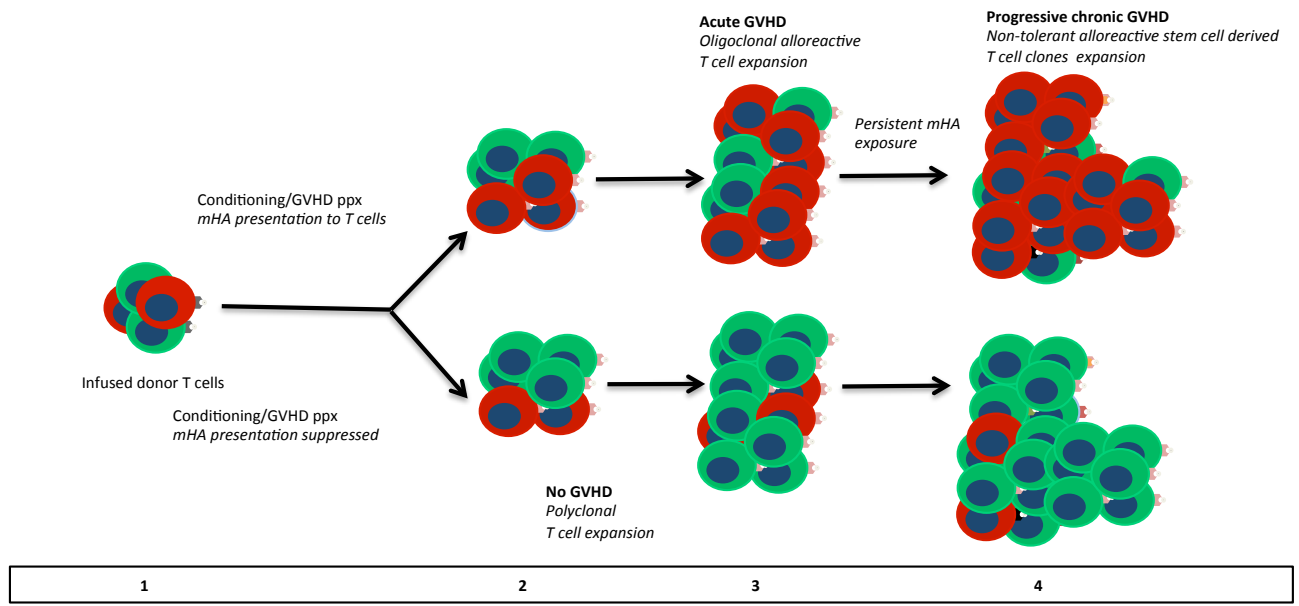

B.

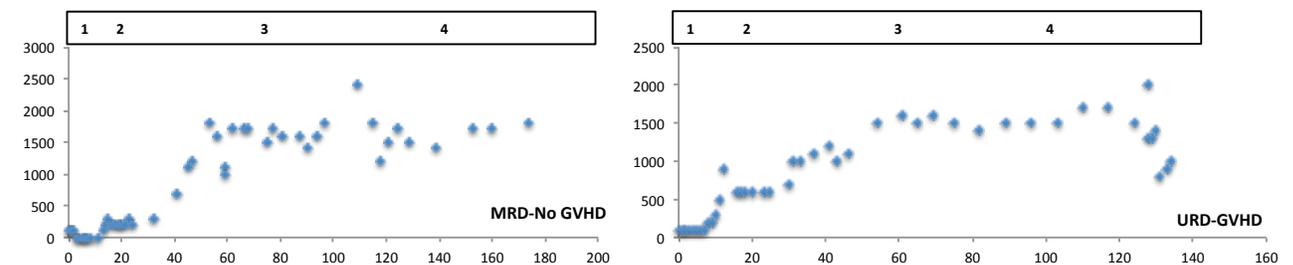

**Figure 4.** Absolute lymphocyte count (ALC, µL⁻¹) plotted as a function of time (days following transplant). (A) Schema of the transplant protocol, outlining the general immunosuppression withdrawal scheme. (B) ALC in the first month following SCT, shows the first growth phase coincident with engraftment. (C) ALC in the second and third month following SCT, shows the second growth phase following cessation of MMF, of these patients only patient D developed GVHD. (D) ALC in the first four to six month following SCT, shows the overall growth kinetics of lymphocytes. Data in all these plots may be modeled with a logistic equation of the general form, $N_t = K/(1+Ae^{-rt})$, where $A=(K-N_0)/N_0$, with $N_0$ represents the lymphocyte count at the beginning, and $Nt$ is the lymphocyte count at time $t$ following transplant, $e$ is the base of natural logarithms, 2.718 and $r$ is the growth rate of the population. A similar equation, $N_t = N_0 +(K-N_0)/(1+10^{(a-t)r})$, where $a$, is the time at which growth rate is maximal and an inflection point is observed in the logistic curve, also describes the data. (E) Patients with clinical events, depicting impact of immunosuppressive therapy on lymphocyte counts, and departure from the sigmoid growth patterns. Also seen is the variability from measurement to measurement in ALC in the fourth month, when comparing patients with GVHD (AA and D) and those without (CC and DD).

**A.**

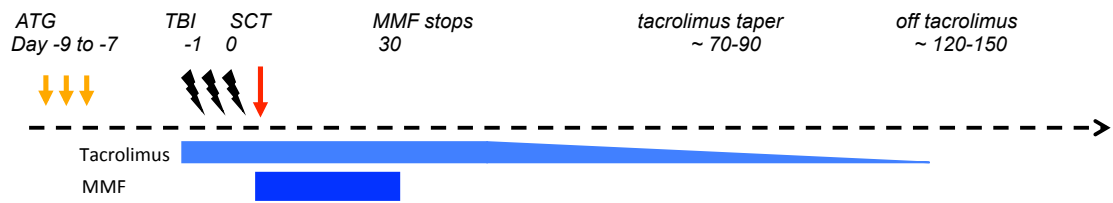

**B.**

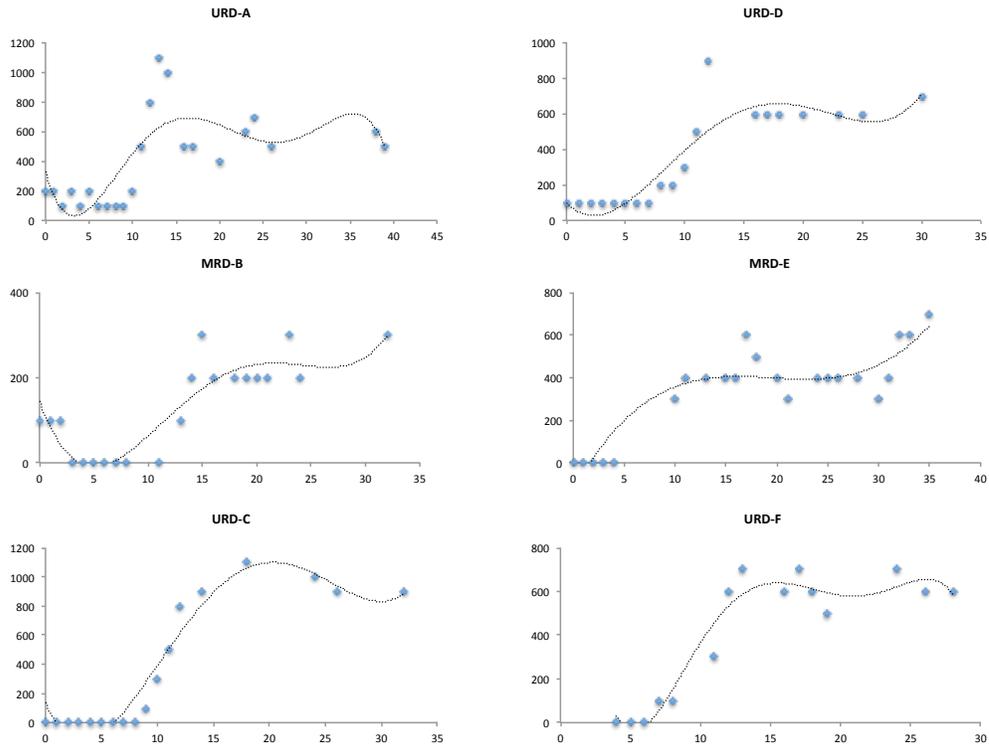

**C.**

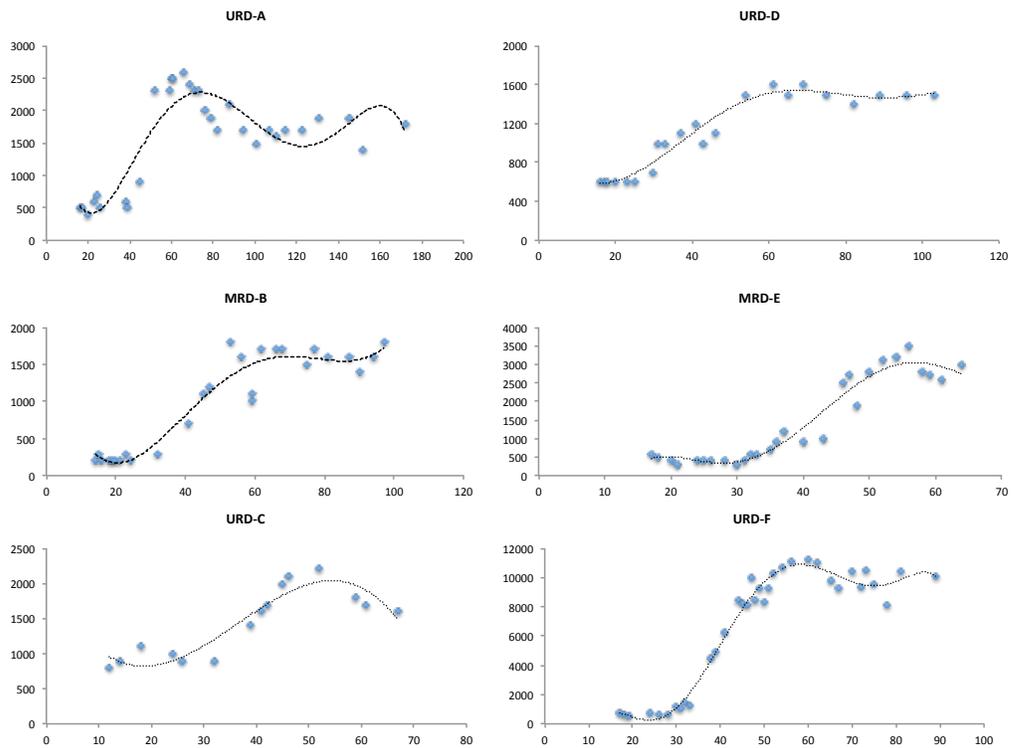

**D.**

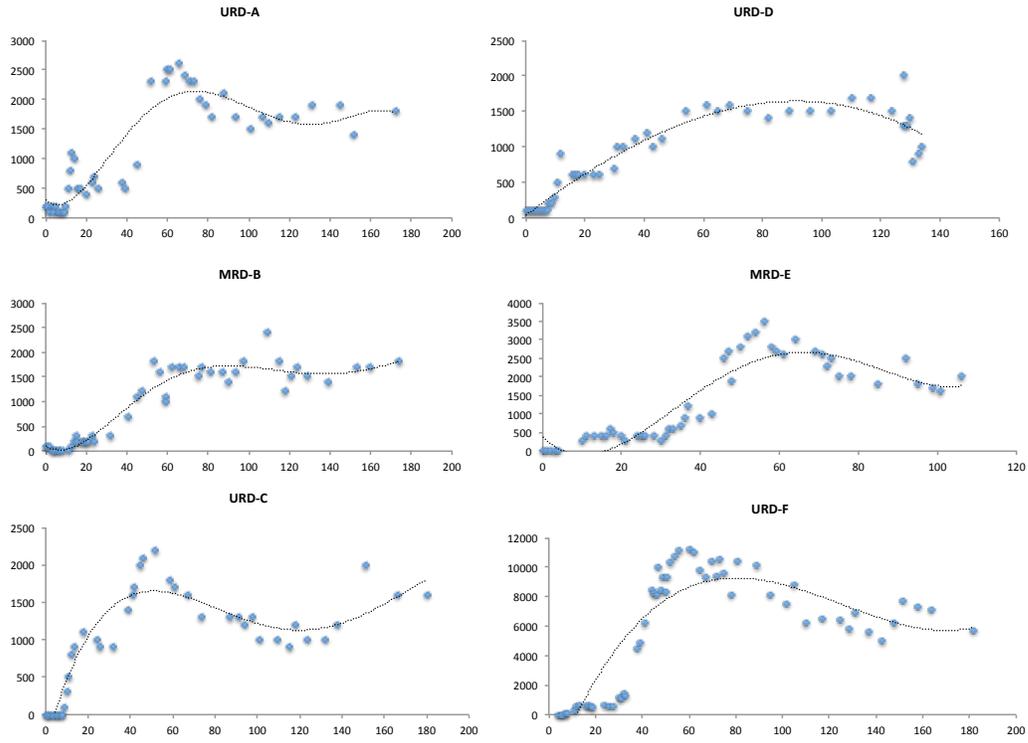

**E.**

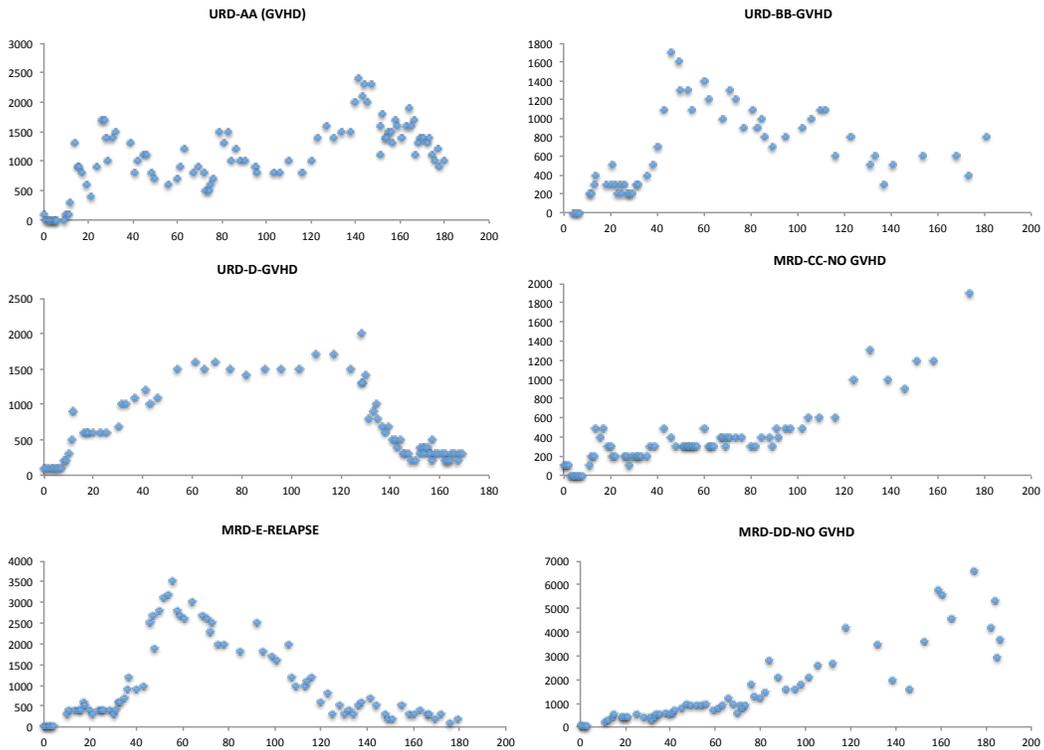

**Appendix**

Modeling stem cell transplantation as a dynamical system. An analogy with a system based in physics will help illustrate this point more clearly, where streams of elementary particles and their secondary emissions are influenced by the electromagnetic fields that they travel in, through time. Transplantation with donor-derived T cells, responding over time to alloreactivity potential (recipient immunogenic mHA-HLA) under the influence of conditioning and immunosuppressive therapy may be similarly considered. Considering a cathode ray tube apparatus, the stream of electrons (e) symbolizes *donor T cells*, the electrical field (E) deflecting the electrons, depicts the *alloreactivity potential* (a function of the immunogenic recipient peptide-HLA complexes that donor T cells encounter), the magnetic field (H) represents the *conditioning and immunosuppressive therapy* that a patient undergoes. The Target is analogous to *tissues* initially encountered by donor T cells, with secondarily emitted electrons (e') illustrating the *donor T cell proliferation*, upon encountering *tissue specific alloreactivity potential* (E') under the influence of diminished *immunosuppression* (H') later in the course of transplantation. Finally, the Signal denotes the *clinical outcome* observed, and like the signal intensity versus location may be normally distributed, with graft loss and fatal GVHD representing the extremes and range of tolerance and milder forms of GVHD, making up the middle.

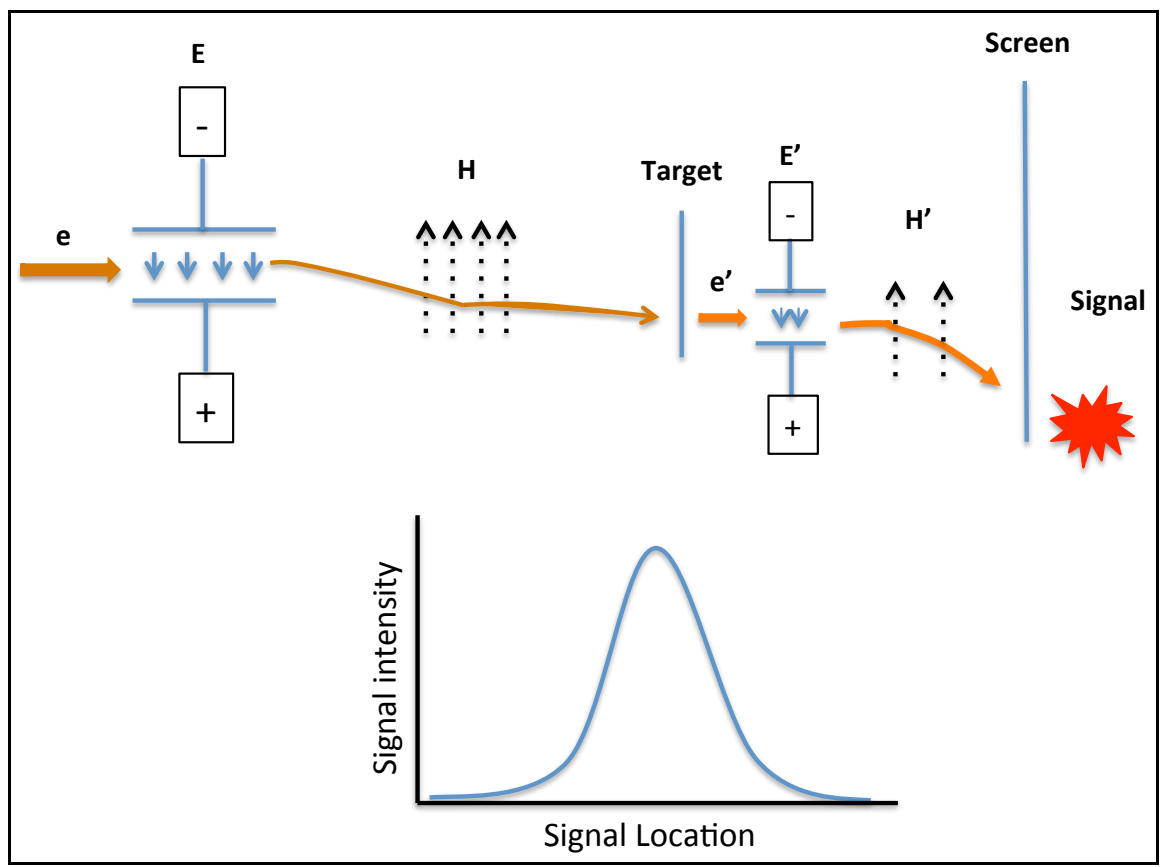

## References:

1.   Horowitz MM, Gale RP, Sondel PM, Goldman JM, Kersey J, Kolb HJ, et al. Graft-versus-leukemia reactions after bone marrow transplantation. *Blood* (1990) **75**:555-62.

2.   Kolb HJ, Schattenberg A, Goldman JM, Hertenstein B, Jacobsen N, Arcese W, et al. Graft-versus-leukemia effect of donor lymphocyte transfusions in marrow grafted patients. *Blood* (1995) **86**:2041-50.

3.   Lee SJ, Klein J, Haagenson M, Baxter-Lowe LA, Confer DL, Eapen M, et al. High-resolution donor-recipient HLA matching contributes to the success of unrelated donor marrow transplantation.
*Blood* (2007) **110**:4576-83.

4.   Valcárcel D, Sierra J, Wang T, Kan F, Gupta V, Hale GA, et al. One-antigen mismatched related versus
HLA-matched unrelated donor hematopoietic stem cell transplantation in adults with acute leukemia: Center for International Blood and Marrow Transplant Research results in the era of molecular HLA typing.
*Biol Blood Marrow Transplant* (2011) **17**:640-8.

5.   Arora M, Weisdorf DJ, Spellman SR, Haagenson MD, Klein JP, Hurley CK, et al. HLA-identical sibling compared with 8/8 matched and mismatched unrelated donor bone marrow transplant for chronic phase chronic myeloid leukemia. *J Clin Oncol* (2009) **27**:1644-52.

6.   Weisdorf DJ, Nelson G, Lee SJ, Haagenson M, Spellman S, Antin JH, et al. Sibling versus unrelated donor allogeneic hematopoietic cell transplantation for chronic myelogenous leukemia: refined HLA matching reveals more graft-versus-host disease but not less relapse. *Biol Blood Marrow Transplant* (2009) **15**:1475-8.

7.   Socié G, Loiseau P, Tamouza R, Janin A, Busson M, Gluckman E, et al. Both genetic and clinical factors predict the development of graft-versus-host disease after allogeneic hematopoietic stem cell transplantation. *Transplantation* (2001) **72**:699-706.

8.   Shlomchik WD. Graft-versus-host disease. *Nat Rev Immunol.* (2007) **7**:340-352.

9.   Thiant S, Yakoub-Agha I, Magro L, Trauet J, Coiteux V, Jouet JP, et al. Plasma levels of IL-7 and IL-15 in the first month after myeloablative BMT are predictive biomarkers of both acute GVHD and relapse.
*Bone Marrow Transplant* (2010) **45**:1546-52.

10. Sivula J, Turpeinen H, Volin L, Partanen J. Association of IL-10 and IL-10Rbeta gene polymorphisms with graft-versus-host disease after haematopoietic stem cell transplantation from an HLA-identical sibling donor. *BMC Immunol* (2009) **10**:24.

11. Schneidawind D, Pierini A, Negrin RS. Regulatory T cells and natural killer T cells for modulation of GVHD following allogeneic hematopoietic celltransplantation. *Blood* (2013) **122**:3116-21.

11. Fleischhauer K, Shaw BE, Gooley T, Malkki M, Bardy P, Bignon JD, et al.  Effect of T-cell-epitope matching at HLA-DPB1 in recipients of unrelated-donor haemopoietic-cell transplantation: a retrospective study.
*Lancet Oncol* (2012) **13**:366-74.

13. Weisdorf D, Spellman S, Haagenson M, Horowitz M, Lee S, Anasetti C, et al. Classification of HLA-matching for retrospective analysis of unrelated donor transplantation: revised definitions to predict survival.
*Biol Blood Marrow Transplant* (2008)**14**:748-58.

11. Barker JN, Weisdorf DJ, DeFor TE, Blazar BR, McGlave PB, Miller JS, et al. Transplantation of 2 partially HLA-matched umbilical cord blood units to enhance engraftment in adults with hematologic malignancy.
*Blood* (2005) **105**:1343-7.

11. Scaradavou A, Brunstein CG, Eapen M, Le-Rademacher J, Barker JN, Chao N, et al. Double unit grafts successfully extend the application of umbilical cord blood transplantation in adults with acute leukemia.
*Blood* (2013) **121**:752-8.

11. Luznik L, O'Donnell PV, Symons HJ, Chen AR, Leffell MS, Zahurak M, et al. HLA-haploidentical bone marrow transplantation for hematologic malignancies using nonmyeloablative conditioning and high-dose,
post-transplantation cyclophosphamide. *Biol Blood Marrow Transplant* (2008)**14**:641-50.

11. Brunstein CG, Fuchs EJ, Carter SL, Karanes C, Costa LJ, Wu J, Devine SM, et al. Alternative donor
transplantation after reduced intensity conditioning: results of parallel phase 2 trials using partially HLA-mismatched related bone marrow or unrelated double umbilical cord blood grafts.
*Blood* (2011) **118**:282-8.

11. Anasetti C. The ever elusive permissive mismatch. *Biol Blood Marrow Transplant* (2012)

**18**:657-8.

11. Spellman S, Klein J, Haagenson M, Askar M, Baxter-Lowe LA, He J, et al. Scoring HLA class I mismatches by HistoCheck does not predict clinical outcome in unrelated hematopoietic stem cell transplantation.
*Biol Blood Marrow Transplant* (2012) **18**:739-46.

20. Portier DA, Sabo RT, Roberts CH, Fletcher DS, Meier J, Clark WB, et al. Anti-thymocyte globulin for
conditioning in matched unrelated donor hematopoietic cell transplantation provides comparable outcomes to matched related donor recipients. *Bone Marrow Transplant* (2012) **47**:1513-9.

21. Mohty M, Labopin M, Balère ML, Socié G, Milpied N, Tabrizi R, et al. Antithymocyte globulins and chronic graft-vs-host disease after myeloablative allogeneic stem cell transplantation from HLA-matched
unrelated donors: a report from the Sociéte Française de Greffe de Moelle et de Thérapie Cellulaire.
*Leukemia* (2010) **24**:1867-74.

22. Koreth J, Stevenson KE, Kim HT, McDonough SM, Bindra B, Armand P, et al. Bortezomib-based
graft-versus-host disease prophylaxis in HLA-mismatched unrelated donor transplantation.
*J Clin Oncol* (2012) **30**:3202-8.

23. Thorn R, Meier J, Wang E, Sabo R, Scalora A, Roberts C, et al. Favorable T Cell Reconstitution In Reduced Intensity Conditioned Allogeneic Stem Cell Transplantation with Low-Dose Rabbit Anti-Thymocyte Globulin and Total Body Irradiation. *Blood* (2013)**122**:4577.

24. Kobulnicky D, Sabo R, Scalora A, Portier D, Fletcher D, Tessier J, et al. Immune Reconstitution in Anti-Thymocyte Globulin Conditioned Unrelated Donor Stem Cell Transplantation. *Blood* (2013) **122**:2071.

25. Toor AA, Sabo RT, Chung HM, Roberts C, Manjili RH, Song S, et al. Favorable outcomes in patients with high donor-derived T cell count after in vivo T cell-depleted reduced-intensity allogeneic stem cell transplantation. *Biol Blood Marrow Transplant* (2012) **18**:794-804.

26. Chiesa R, Gilmour K, Qasim W, Adams S, Worth AJ, Zhan H, et al. Omission of in vivo T-cell depletion promotes rapid expansion of naïve CD4+ cord blood lymphocytes and restores adaptive immunity within 2 months after unrelated cord blood transplant. *Br J Haematol* (2012) **156**:656-66.

27. Barrett J. Improving outcome of allogeneic stem cell transplantation by immunomodulation of the early post-transplant environment. *Curr Opin Immunol* (2006) **18**:592-8.

28. Savani BN, Mielke S, Rezvani K, Montero A, Yong AS, Wish L, et al. Absolute lymphocyte count on day 30
is a surrogate for robust hematopoietic recovery and strongly predicts outcome after T cell-depleted allogeneic stem cell transplantation. *Biol Blood Marrow Transplant* (2007) **13**:1216-23.

29. Pavletic ZS, Joshi SS, Pirruccello SJ, Tarantolo SR, Kollath J, Reed EC, et al. Lymphocyte reconstitution after allogeneic blood stem cell transplantation for hematologic malignancies. *Bone Marrow Transplant* (1998) **21**:33-41.

30. Bartelink IH, Belitser SV, Knibbe CA, Danhof M, de Pagter AJ, Egberts TC, et al. Immune reconstitution kinetics as an early predictor for mortality using various hematopoietic stem cell sources in children.
*Biol Blood Marrow Transplant* (2013) **19**:305-13.

31. Giannelli R, Bulleri M, Menconi M, Casazza G, Focosi D, Bernasconi S, et al. Reconstitution rate of absolute CD8+ T lymphocyte counts affects overall survival after pediatric allogeneic hematopoietic stem cell transplantation. *J Pediatr Hematol Oncol* (2012) **34**:29-34.

32. Sampson J, Sheth N, Koparde V, Scalora A, Serrano M, Lee V, Whole Exome Sequencing to Estimate Alloreactivity Potential Between Donors and Recipients in Stem Cell Transplantation. *Blood* (2013) **122**:150.

33. Meier J, Roberts C, Avent K, Hazlett A, Berrie J, Payne K, et al. Fractal organization of the human T cell repertoire in health and after stem cell transplantation. *Biol Blood Marrow Transplant* (2013) **19**:366-77.

34. Grebogi C, Ott E, Yorke JA. Chaos, strange attractors, and fractal basin boundaries in nonlinear dynamics. *Science* (1987) **238**:632-8.

35. Lorenz, Edward N. Deterministic Nonperiodic Flow. *Journal of Atmospheric Science* **20**(1963):130-141.

36. Stewart I. "The Imbalance of Nature Chaos Theory,". In: *In Pursuit Of The Unknown, 17 Equations That Changed The World*. Philadelphia, PA: Basic Books (2012). p. 283.

37. Liebovitch, L. S. and Scheurle, D. Two lessons from fractals and chaos. Complexity (2000) **5**: 34–43.

38. Baron F, Petersdorf EW, Gooley T, Sandmaier BM, Malkki M, Chauncey TR, et al. What is the role for donor natural killer cells after nonmyeloablative conditioning? *Biol Blood Marrow Transplant* (2009) **15**:580-8.

39. Jameson-Lee M, Koparde VN, Sampson JK, Scalora AF, Khalid H, Sheth N. In Silico Derivation of HLA-Specific Alloreactivity Potential from Whole Exome Sequencing of Stem Cell Transplant Donor-Recipient Pairs. *Biol Blood Marrow Transplant* (2014) **20**:S269-70. *Also see*, Jameson-Lee *et al*. In Silico Derivation of HLA-Specific Alloreactivity Potential from Whole Exome Sequencing of Stem Cell Transplant Donors and Recipients: Understanding the Quantitative Immuno-biology of Allogeneic Transplantation. arXiv:1406.3762v1.

40. May RM. Biological populations with nonoverlapping generations: stable points, stable cycles, and chaos. *Science* (1974) **186**:645-7.

41. Freeman JD, Warren RL, Webb JR, Nelson BH, Holt RA. Profiling the T-cell receptor beta-chain repertoire by massively parallel sequencing. *Genome Res* (2009) **19:**1817-24.

42. Neale MC, McArdle JJ. Structured latent growth curves for twin data. *Twin Research* (2000) **3**:165–177

43. Waliszewski P, Konarski J. "A Mystery of the Gompertz Function,". In: Losa GA Editor. *Fractals in Biology and Medicine, Vol IV.* Book Series: Mathematics and Biosciences in Interaction (2005). p. 277-86.

44. Bellucci R, Alyea EP, Weller E, Chillemi A, Hochberg E, Wu CJ, et al. Immunologic effects of prophylactic donor lymphocyte infusion after allogeneic marrow transplantation for multiple myeloma.
*Blood* (2002) **99**:4610-7.

45. Berrie JL, Kmieciak M, Sabo RT, Roberts CH, Idowu MO, Mallory K, et al. Distinct oligoclonal T cells are associated with graft versus host disease after stem-cell transplantation.*Transplantation* (2012) **93**:949-57.

46. Tsutsumi Y, Tanaka J, Miura Y, Toubai T, Kato N, Fujisaw F, et al. Molecular analysis of T-cell repertoire in patients with graft-versus-host disease after allogeneic stem cell transplantation. *Leuk Lymphoma* (2004) **45**:481-8.

47. Liu C, He M, Rooney B, Kepler TB, Chao NJ. Longitudinal analysis of T-cell receptor variable beta chain repertoire in patients with acute graft-versus-host disease after allogeneic stem cell transplantation. *Biol Blood Marrow Transplant* (2006) **12**:335-45.

48. Luca Gattinoni, Daniel J. Powell, Jr., Steven A. Rosenberg & Nicholas P. Restifo. Adoptive immunotherapy for cancer: building on success. *Nature Reviews Immunology* (2006) 6; 383-393